\documentclass[11pt,twoside]{article}

\usepackage{asp2006}
\usepackage{epsf}
\usepackage{lscape}
\usepackage{graphicx}

\begin{document}
\title{Spectropolarimetry Surveys of Obscured AGNs}
\author{Edward C.\ Moran}
\affil{Astronomy Department, Wesleyan University, Middletown, CT 06459 USA}

\begin{abstract}
The results of spectropolarimetry surveys of obscured AGNs are reviewed,
paying special attention to their implications for the existence of two
populations of type~2 Seyfert galaxies --- hidden Seyfert~1s and ``true''
Seyfert~2s.  In this context, the results from our Keck spectropolarimetry
survey are presented.  Similar to previous work, we have detected hidden
broad-line regions (HBLRs) in about half of our sample.  However, owing
to different selection methods, we find that differences in the properties
of HBLR and non-HBLR objects are much less pronounced than prior
reports have indicated.  Spectropolarimetry studies continue to benefit
AGN research, as illustrated by the discovery of double-peaked
H$\alpha$ emission in the polarized-flux spectrum of NGC~2110.

\end{abstract}

\section{Introduction}
Spectropolarimetry allows us to measure the polarization of light as a
function of wavelength.  Coupled with theory and other observational
constraints, this information can reveal whether the origin of the detected
polarization is intrinsic (as in the case of synchrotron emission) or
extrinsic, i.e., imparted via transmission through or scattering by a
medium.  When scattering is the dominant mechanism, spectropolarimetric
observations can provide insight into the structure and geometry of the
emitting region that would be unobtainable by other means.  It is this
aspect of the technique that has had such a profound impact on AGN research.

In keeping with the ``torus'' theme this group of papers, I'll limit this
review to spectropolarimetry surveys of obscured AGNs, which locally
are classified as type~2 Seyfert galaxies on the basis of their narrow
emission-line optical spectra.  Contrary to the stated aims of this
meeting, my focus is more on the astronomical issues associated with
spectropolarimetry surveys, rather than on the physics of AGNs.  However,
this is worthwhile, since a number of recent investigations into the physical
nature of AGNs have been motivated by spectropolarimetry results.

\section{Pioneering Work at Lick Observatory}

By now, the basic spectropolarimetric results obtained for Seyfert~2 galaxies
are familiar to most.  Beginning with NGC~1068 (Miller \& Antonucci 1983;
Antonucci \& Miller 1985; Miller, Goodrich, \& Matthews 1991), it has been
demonstrated that some Seyfert~2s display broad permitted emission lines
in their polarized-light spectra.  In these cases, the data suggest that 
the objects are in fact type~1 Seyfert nuclei whose continuum source and
broad-line region are hidden from direct view by dense material.  Emission
from the nucleus emerges along another direction and is scattered into
our line-of-sight by free electrons or dust.  Because the scattered light is
polarized, spectropolarimetry can provide us with a periscopic view of the
innermost regions of an obscured active nucleus.  With support from some
additional Lick Observatory discoveries of hidden Seyfert~1s (Miller \&
Goodrich 1990; Tran, Miller, \& Kay 1992), this picture provided an elegant
way to unify the two major spectroscopic classes of AGNs.

Spectropolarimetry measurements impose constraints on the geometry of the
obscuring medium in AGNs.  To account for the high intrinsic polarizations
of some Seyfert~2s (tens of percent; Tran 1995), the nuclear radiation field
must be restricted in both the plane of scattering (scattering angles must
be close to $\sim 90^{\circ}$) and in the plane of the sky (position angles
of the scattering planes, as viewed by us, must be similar to obtain net
polarization).  Thus, the obscuration is such that it surrounds the nucleus
and has some cylindrical symmetry --- a torus, more or less.  Support for
the torus model was derived early on from (1) the fact that the polarization
position angle tends to be orthogonal to the axis of the nuclear radio source
in type~2 AGNs (Antonucci 1984), and (2) the discovery of optical ionization
cones in the circumnuclear regions of some Seyfert~2 galaxies (e.g., Pogge
1989), which demonstrate directly that the nuclear radiation field is
anisotropic.  Spectropolarimetry also provides an indication of where the
torus is located.  When the data are good enough, we observe that the narrow
emission lines are unpolarized, which implies that most of the scattering
occurs interior to the narrow-line region.  Thus, the obscuration must be
located within a few parsecs of the black hole.

For her Ph.D.\ thesis, L.~Kay (1990, 1994) performed what was the first
substantial spectropolarimetry survey of Seyfert~2 galaxies.  Observations
of 50 objects were made in the blue to explore their continuum polarizations
at wavelengths where dilution from host-galaxy starlight is reduced.  Her
study demonstrated that ``raw'' continuum polarizations are uniformly low,
and that the fraction of the continuum associated with unpolarized bulge
starlight --- even within the small apertures used --- tends to be very high.
Host-galaxy dilution remains an important factor for spectropolarimetry surveys
performed at H$\alpha$.

\section{Spectropolarimetry Surveys at H$\alpha$}

The first extensive search for polarized broad H$\alpha$ emission in
narrow-line AGNs was carried out by Young et al.\ (1996).  Using the AAT
3.9-m telescope, they surveyed a sample of 24 galaxies consisting of
well-known Seyfert~2s and {\sl IRAS\/} galaxies with ``warm''
25~$\mu$m/60~$\mu$m infrared colors.  Although the Young et al.\ sample
was not complete or unbiased, their results would foreshadow those of
spectropolarimetry surveys to come:\ a number of the objects exhibit
polarized broad lines, but the majority do not.

Combining data from the AAT 3.9-m and detections of hidden broad-line regions
(HBLRs) from the literature, Heisler, Lumsden, \& Bailey (1997) surveyed a
small but statistically complete sample of 16 objects assembled on the basis
of their far-infrared fluxes and luminosities.  The discovery of a new HBLR
increased the number of objects in the sample with polarized
broad lines to 7, or 44\%.  Using similar but somewhat looser selection
criteria, Lumsden et al.\ (2001) examined an expanded {\sl IRAS\/} galaxy
sample of 24 narrow-line AGNs.  Again, 4-m class telescopes were employed.
Only one new detection was reported, indicating that 33\% (8 objects) in
the larger sample have evidence for polarized broad lines.

Seeking to improve upon the statistical quality of previous surveys, Tran
(2001, 2003) studied 49 Seyfert~2 galaxies from the CfA and 12~$\mu$m AGN
samples.  Five new HBLRs were reported among the objects, which were observed
with the Lick 3-m and Palomar 5-m telescopes.  The total number of objects
in the Tran survey with hidden broad-line regions (22, or 45\% of the sample)
allowed for detailed comparisons between object that do and do not have
evidence for polarized broad emission lines.

The survey of Seyfert~2 galaxies carried out by me and my colleagues is
based mainly on data obtained with the Keck 10-m telescopes (Moran et al.\
2000, 2001, 2007).  In addition, the sample we have used is significantly
different than those employed by other investigators; it consists of 38
nearby objects from the distance-limited sample of Ulvestad \& Wilson (1989,
hereafter UW89).
%Most of the objects are bona fide type~2 galaxies, though 4 ``narrow-line
%X-ray galaxies'' in the sample are technically Seyfert~1.9s.
The 9 new HBLR objects we have discovered among the UW89 galaxies brings to
17 the total number detected to date, which translates to 45\% of the sample.
Thus, the statistical power of our survey is on par with that of the Tran
study.

Though not strictly a Seyfert~2 survey, I want to point out the results of
A.~Barth's Ph.D.\ thesis, in which he explored the polarization properties
of 14 low-luminosity AGNs.  The objects, drawn from the spectroscopic survey
of Ho, Filippenko, \& Sargent (1995), included a few Seyfert~2s and Seyfert~1s,
and several LINERs.  Remarkably, the data revealed that three of the LINERs
(located in elliptical galaxies) have polarized broad emission lines (Barth,
Filippenko, \& Moran 1999), implying that they possess nuclear obscuration
with a geometry similar to that present in Seyfert galaxies.

\section{Polarization and the Nature of Seyfert~2 Galaxies}

Spectropolarimetry surveys have indicated that at least $\sim$~half of all
Seyfert~2 galaxies have broad-line regions that are obscured from direct view
by dense nuclear material.  Are the non-HBLR objects hidden Seyfert~1s for
which the detection of polarization is especially challenging (caused by
something physically uninteresting, e.g., an unfavorable scattering geometry),
or do they differ from hidden Seyfert~1s in some fundamental way?  This is
an important question --- it concerns a significant fraction of all Seyfert
galaxies, and as such, it has implications for our overall understanding of the
AGN phenomenon and other associated issues, e.g., the origin of the cosmic
X-ray background.

One of the first attempts to compare the properties of Seyfert~2s that do
and do not have evidence for polarized broad lines was presented in my
first paper (Moran et al.\ 1992).  We reported that the nuclear radio
luminosities of the nine hidden Seyfert~1s known at that time are all high
compared to the general Seyfert population.  This suggested that perhaps
there are two types of Seyfert~2s --- hidden Seyfert~1s and ``true''
Seyfert~2s.  Current investigations center around luminosity differences
as well, and as with our early attempt, it's the {\it interpretation\/} of
those differences that presents the main challenge.

\subsection{Comparing Properties: HBLRs vs.\ Non-HBLRs}

A number of authors have compared the properties of HBLR and non-HBLR Seyfert
2s.  These include those who have conducted spectropolarimetry studies, as
well as those who have performed larger meta-analyses based on collections
of results from the literature (e.g., see papers by Q.~Gu and collaborators).
The literature samples afford greater statistical power, although systematic
effects or biases in the original samples will be reinforced (if not amplified)
under this approach.

The radio luminosities of HBLR and non-HBLR Seyfert~2s apparently do
differ, though not to the extent that we originally noted.  I should point
out that some of the comparisons have been made using total (or nearly so)
radio flux densities obtained from the NVSS.  The radio luminosities of
Seyfert nuclei are often comparable to those of their host galaxies, so
using the total emission can hinder the comparison when the nucleus does
not dominate.  In terms of total radio emission, Lumsden et al.\ (2001)
found no evidence for HBLR Seyfert 2s to be more powerful radio sources
than non-HBLRs, while Gu \& Huang (2002) and Tran (2003) report that HBLRs
tend to be significantly stronger sources.  Thean et al.\ (2001) found that
core radio luminosities are higher on average among HBLRs; only a marginal
difference was observed in the Lumsden et al.\ (2001) study.

The luminosity of the [O~III] $\lambda$5007 line represents an isotropic
indicator of the nuclear luminosity in an AGN, provided that the narrow-line
regions of objects are similar (e.g., in terms of density and covering
fraction), that the spectroscopic data have been acquired under photometric
conditions, and that other factors (e.g., slit losses and internal
reddening corrections) are not major sources of additional uncertainty.
All of the above authors have found that HBLRs tend to be more luminous
[O~III] emitters than non-HBLRs, though it should be noted that there is
significant overlap between the two populations.

Similarly, all previous studies have found that HBLRs tend to have warmer
far-infrared colors (i.e., higher $S_{25\mu{\rm m}}/S_{60\mu{\rm m}}$ ratios)
than non-HBLRs.  The effect was first discussed by Heisler et al. (1997),
who reported that the distribution in this parameter is bimodal for HBLRs
and non-HBLRs. Lumsden et al.\ (2001) confirmed the result with their expanded
sample, and strong differences in the FIR colors of HBLRs and non-HBLRs
are present in the CfA/12~$\mu$m sample (Tran 2003) and in the literature
as a whole (Gu \& Huang 2002).  As the FIR emission of AGNs tends to be
warm and that of their host galaxies tends to be cool, we can conclude
that HBLRs contribute a greater fraction of the total flux in the large
{\sl IRAS\/} beam than non-HBLRs do, which suggests that the HBLRs are
intrinsically more luminous.

In summary, spectropolarimetry surveys performed to date have found there
to be significant differences between HBLRs and non-HBLRs in properties
that relate to the luminosity of the nucleus.  Do these differences point
to two types of Seyfert~2s?  Given essentially the same evidence, various
authors have come down on different sides of the debate.  Tran (2001,
2003) has argued that the higher luminosities of the HBLRs in his sample,
coupled with the similarity between HBLRs and Seyfert~1s in comparisons
of orientation-independent parameters, suggest there are two populations
of Seyfert~2 nuclei.  Others (Kay 1994; Lumsden et al.\ 2001; Lumsden \&
Alexander 2001; Gu \& Huang 2002) have countered that, all things being
equal, polarized broad emission lines {\it should\/} be easier to detect
among nuclei of higher luminosity.  In such cases there is greater contrast
between the scattered nuclear component and the unpolarized bulge
starlight in the spectrograph aperture, which should make it easier to
detect a weak polarization signal.  Lumsden et al.\ (2001) demonstrated
the plausibility of this argument by showing that the equivalent width
of the [O~III] line --- which represents the contrast between the nuclear
line emission and the diluting host-galaxy flux --- is higher in HBLRs
than in non-HBLRs.

\subsection{UW89 Results}

Our spectropolarimetry observations of the UW89 Seyfert~2s are valuable
for two reasons.  First, they are sensitive, owing to the large aperture of
Keck and the excellent image quality often obtained on Mauna Kea, which
permits the use of a narrow slit (and the exclusion of much of the unpolarized
bulge starlight).  Second, our sample is distance-limited, and it differs
significantly from the samples used in previous studies.  The
{\sl IRAS}-selected sample of Lumsden et al.\ (2001) and the CfA/12~$\mu$m
sample of Tran (2003) are both flux-limited.  Thus, despite the fact that
they are complete and clearly defined, they must to some degree suffer
from Malmquist effects, i.e., an over-representation of rare, luminous
objects, and an under-representation of the weaker objects that dominate
the population.  A distance-limited sample should more accurately represent
the true distribution of luminosities in a given population.  As Figure~1
illustrates, the far-infrared and 25~$\mu$m luminosities of the CfA/12~$\mu$m
objects are indeed biased to higher values relative to the UW89 sample.
In terms of $L_{25\mu{\rm m}}$, which provides a good measure of the strength
of the nucleus (Lumsden \& Alexander 2001), two-thirds of the CfA/12~$\mu$m
objects are brighter than $10^{10}$~$L_{\odot}$; two-thirds of the UW89
galaxies are {\it less\/} luminous than this.

\begin{figure}[bh]
\centerline{\plottwo{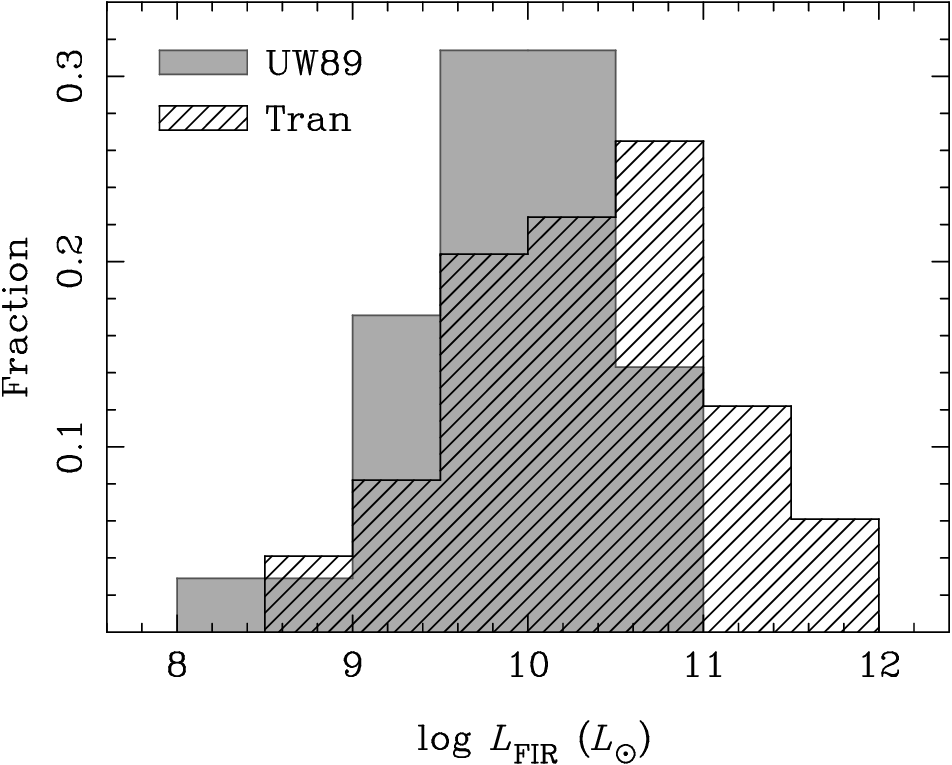}{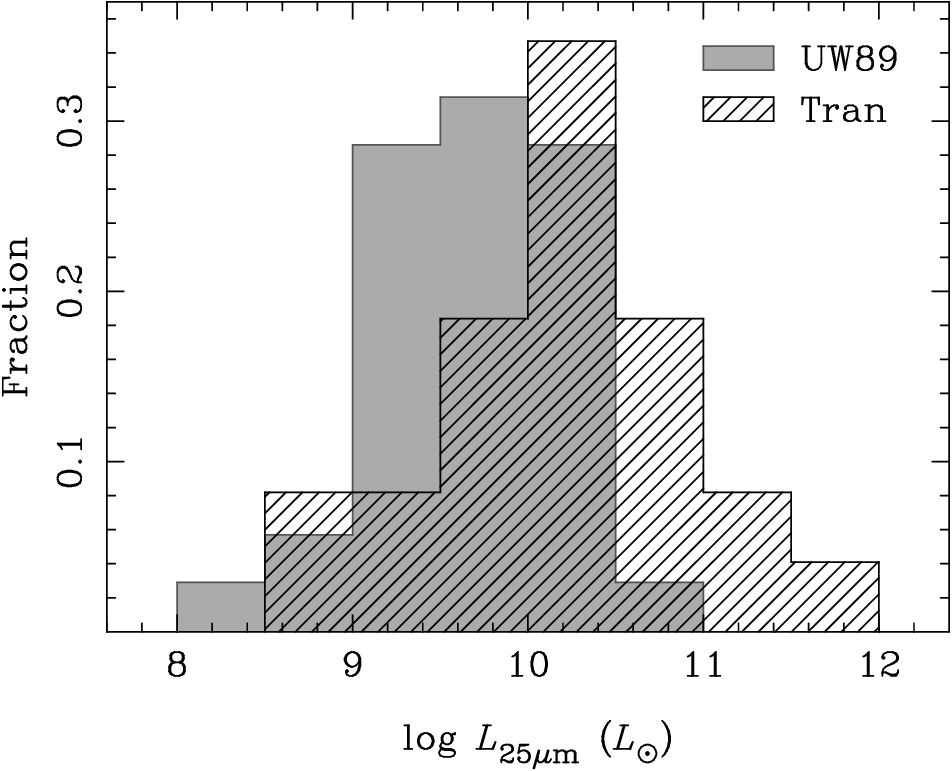}}
\caption{Comparison of the total far-infrared ({\it left}) and 25~$\mu$m
({\it right}) luminosities of the CfA/12~$\mu$m and UW89 samples.}
\end{figure}

A comparison of the nuclear radio and mid-IR luminosities of the HBLRs and
non-HBLRs in the UW89 sample is displayed in Figure~2.  The HBLRs tend to
be more luminous sources, but the differences are not nearly as dramatic as
previous reports have indicated.  And unlike the CfA/12~$\mu$m objects, some
of the UW89 HBLRs are found in genuine low-luminosity objects.  For example,
only 2/22 CfA/12~$\mu$m HBLRs are less luminous than $L_{25\mu{\rm m}} =
10^{10}\, L_{\odot}$, whereas over half of the UW89 HBLRs have
$L_{25\mu{\rm m}} < 10^{10}\, L_{\odot}$.

\begin{figure}[th]
\centerline{\plottwo{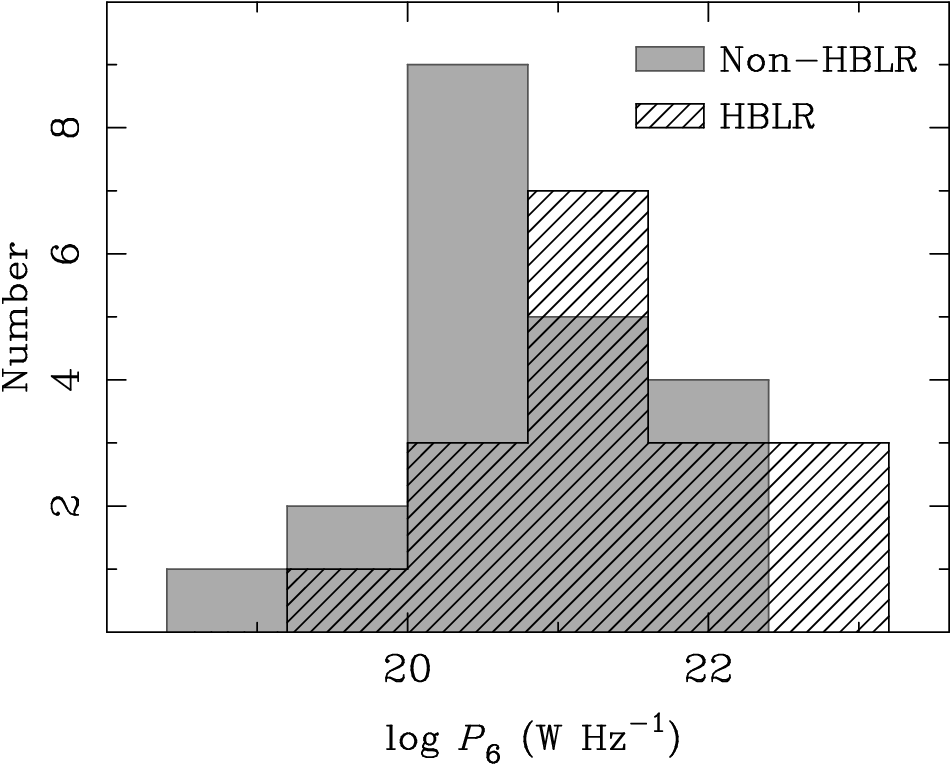}{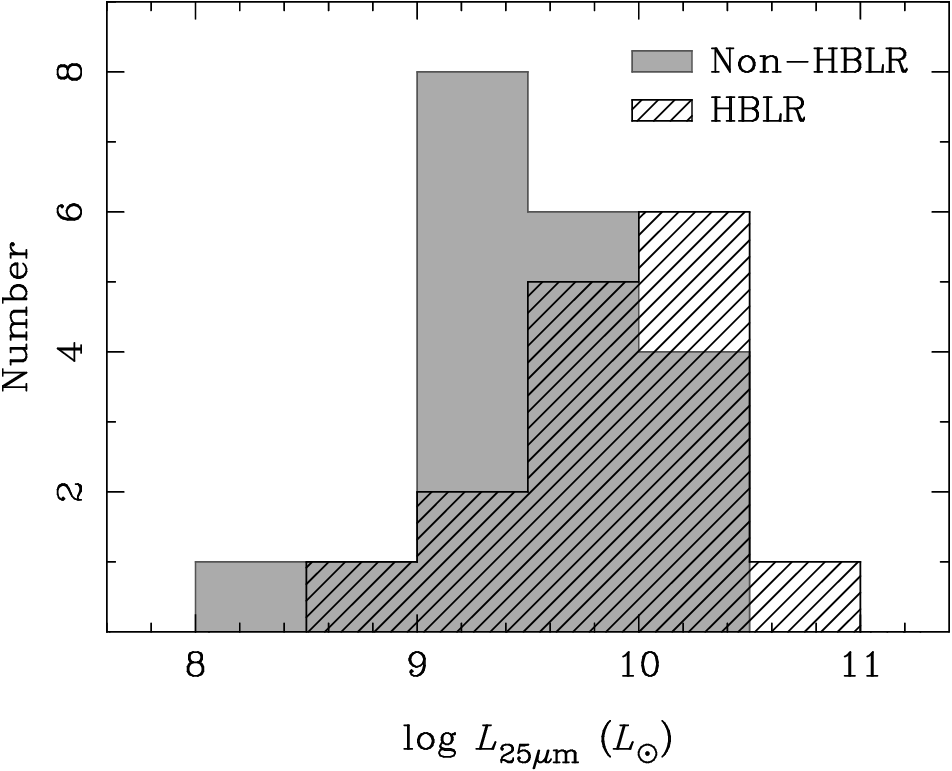}}
\caption{Distribution of the ({\it left}) nuclear 6~cm radio powers and
({\it right}) 25~$\mu$m luminosities for HBLR and non-HBLR objects in the
UW89 sample.}
\end{figure}

We have obtained similar results for the IR
colors of the UW89 Seyfert~2s.  As shown in Figure~3, the HBLRs tend to be
a little warmer, but their $S_{25}/S_{60}$ ratios cover the same range of
values as the non-HBLRs, and the differences in the distributions do not
appear to be that significant.  In contrast, there is a {\it much\/} greater
division of the IR colors of HBLRs and non-HBLRs in the CfA/12~$\mu$m sample.

\begin{figure}[htb]
\centerline{\plottwo{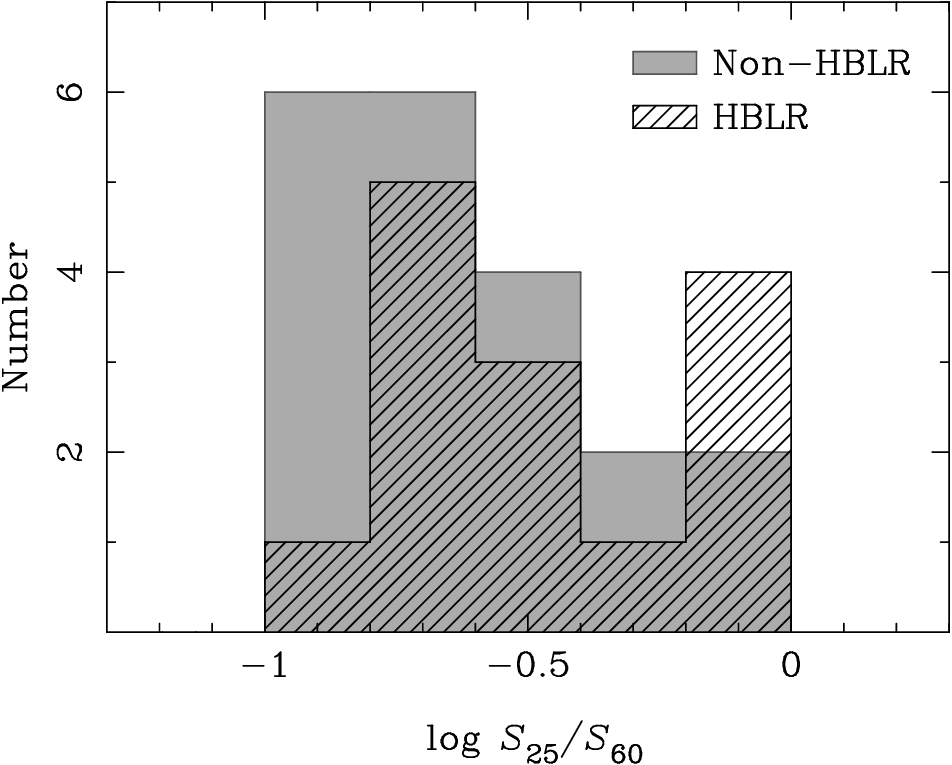}{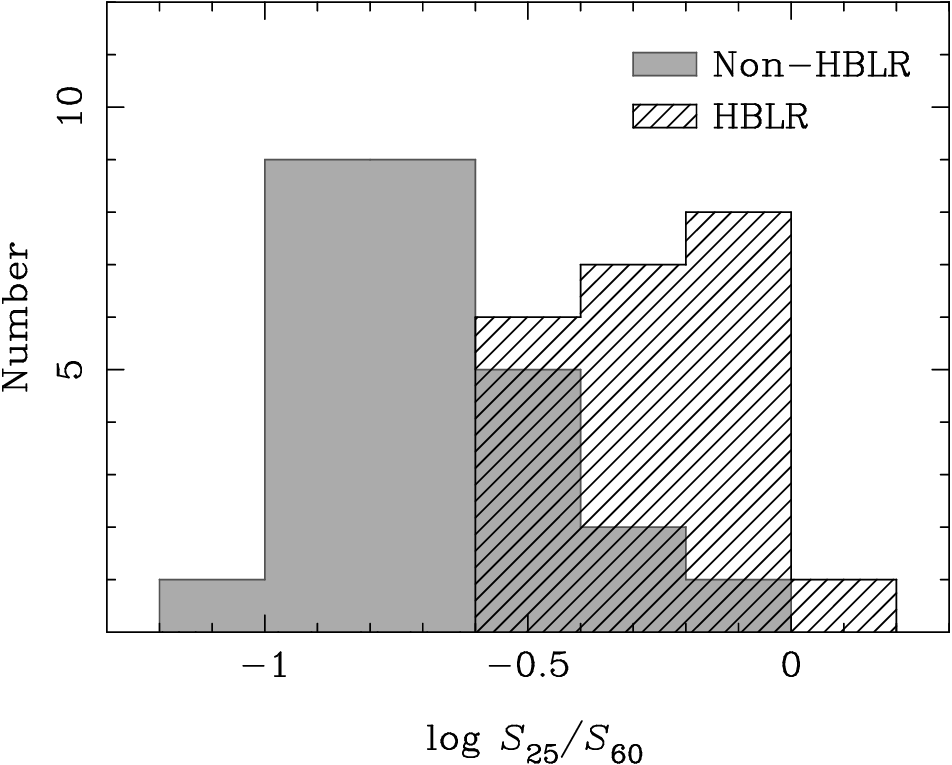}}
\caption{Distribution of the mid-infrared colors of HBLR and non-HBLR
Seyfert~2s in the UW89 ({\it left}) and CfA/12~$\mu$m ({\it right}) samples.}
\end{figure}

In Figure~4, we plot the distribution of [O~III] equivalent widths (EWs)
for the UW89 sample.  The measurements were made directly from our
spectropolarimetry data, so they reflect the contrast between the nucleus
and host galaxy in our observations.  While the objects with the highest EWs
are HBLRs and those with lowest EWs are not, there is a great deal of overlap
between the EW distributions --- far more than that present in the Lumsden
et al.\ (2001) study.

\begin{figure}[htb]
\centerline{\includegraphics[scale=0.7]{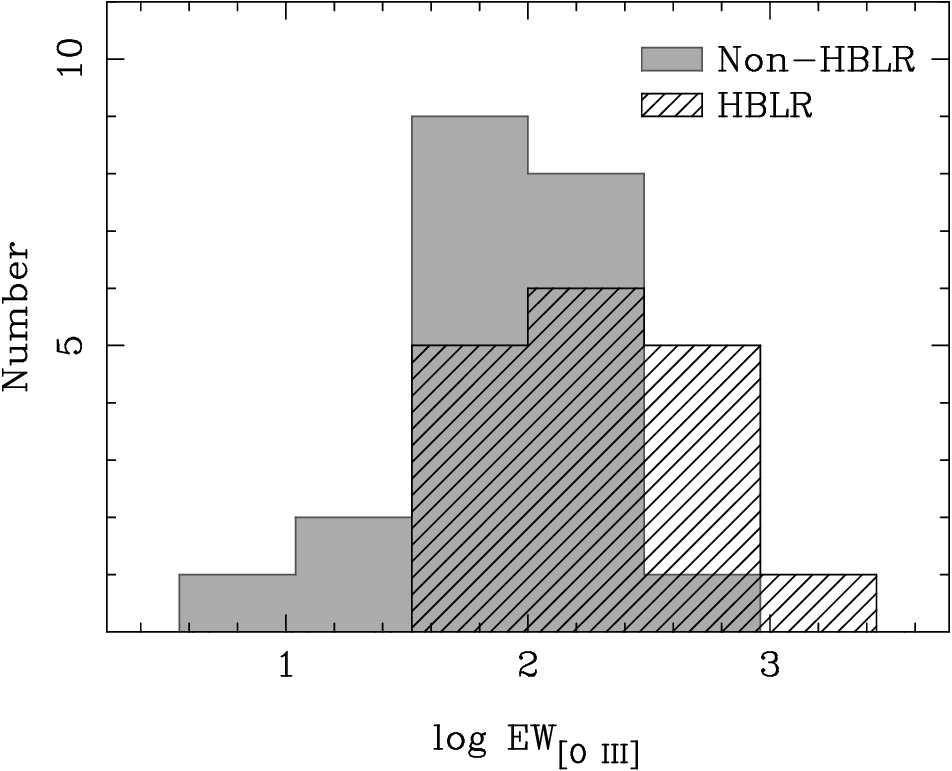}}
\caption{Distribution of the [O III] $\lambda5007$ equivalent widths of the
UW89 Seyfert~2s.}
\end{figure}

Finally, we consider the X-ray properties of the UW89 Seyfert~2s, which have
been discussed by Moran et al.\ (2001) and Cardamone, Moran, \& Kay (2007).
Although the UW89 sample consists of the nearest Seyfert~2s, their observed
luminosities are modest.  As a result, many of the objects were only weakly
detected in long {\sl ASCA\/} observations.  Coupled with the likelihood
that a number of them are Compton-thick, we are not in a position to compare
the absorption column densities or the intrinsic X-ray luminosities of the
HBLRs and non-HBLRs in the sample.  However, we have constructed composite
1--10 keV X-ray spectra for the two groups of objects  (see Fig.~5).
Overall, the composite spectra have the same basic shape, indicating that
the objects in both groups are heavily absorbed sources.  Collectively, the
non-HBLRs are less bright than the HBLRs, and, as evidenced by the large
equivalent width of the Fe~$K\alpha$ line, they are probably more absorbed
on average.

\begin{figure}[htb]
\centerline{\includegraphics[scale=0.52,angle=270]{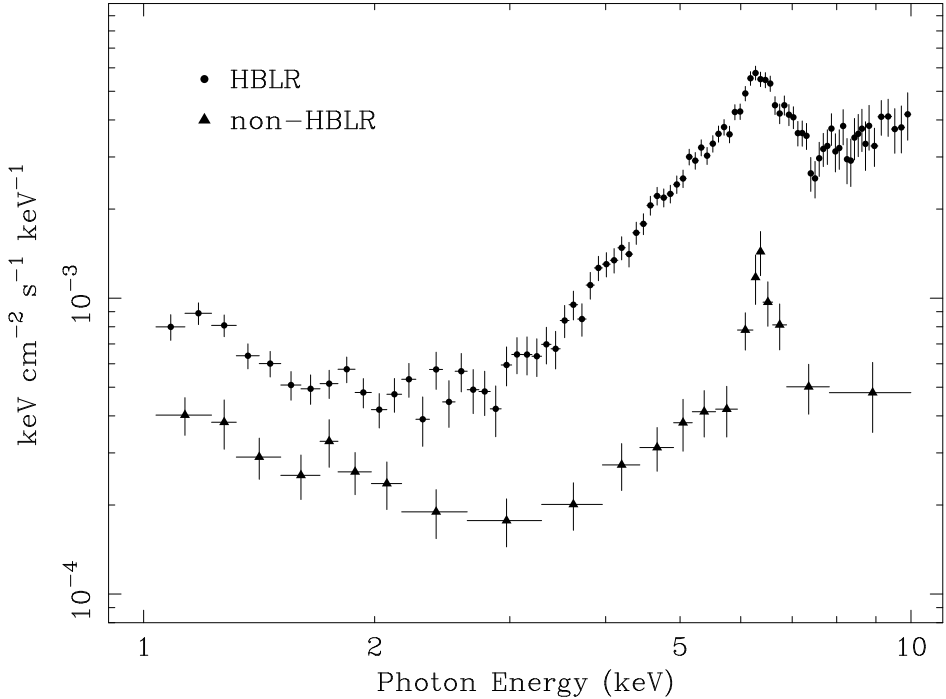}}
\caption{Composite X-ray spectra of the HBLRs and non-HBLRs in the UW89
sample (from Moran et al.\ 2001).  The contributions of the narrow-line
X-ray galaxies in the sample are not included in the composite spectra.}
\end{figure}

\section{What Does It All Mean?}

One of the important lessons we have learned from our survey is that bigger
{\it is\/} better.  Before turning to Keck, we observed much of the UW89
sample with the Lick 3-m for a few hours per galaxy.  Several objects ---
NGC~591, NGC~2273, NGC~5347, and NGC~5929 --- showed nothing in the 3-m data.
But in just 20 minutes apiece with Keck, we were able to detect polarized broad
lines in them.  The moral is: Take care when interpreting spectropolarimetry
non-detections.

On one level, the results of the UW89 survey are consistent with those of
previous studies --- just under half of the objects are HBLRs, and these
tend to be more luminous than the non-HBLRs.  However, we have detected
HBLRs in some intrinsically weak objects whose luminosities fall within
ranges formerly occupied only by non-HBLRs.  In addition, the differences
between the two groups are not that pronounced in our survey.  In particular,
the well-known dichotomy in the IR colors of HBLRs and non-HBLRs is
considerably weaker in the UW89 sample.  Taken together, these results cast
doubt on the existence of a significant population of ``true'' Seyfert~2s.

Overall, our findings support the idea that the luminosity differences we
observe arise (in part) because a weak polarization signal is easier to detect
in objects with brighter nuclei.  But the significant overlap we see in the
properties of HBLRs and non-HBLRs suggests that luminosity is not the only
factor relevant to the detection of polarized broad lines.  If it were, we
would expect a greater separation between HBLRs and non-HBLRs in the [O~III]
equivalent width ``contrast'' parameter.  It seems likely that another
parameter is involved --- perhaps something that affects the amount of
scattered flux we receive from a particular object.  If so, it is probably
unrelated to luminosity, given the lack of strong differences we have found
among the HBLRs and non-HBLRs in the UW89 sample.

\section{Future Prospects}

Spectropolarimetry still has a lot to offer AGN research.  Our Keck study
has demonstrated that the distinction between HBLR and non-HBLR Seyfert~2s
becomes blurred when sensitive data are obtained for a sample that is free
of serious luminosity bias.  The emphasis of future work should be on
improving and enlarging the samples employed.  Completeness at lower
luminosities is essential, which argues for samples of the nearest objects.
And future surveys should be more sensitive --- most of our Keck observations
were made in ``survey mode'' at just 20 minutes per object, which leaves
plenty of room for increased sensitivity.

In addition to more robust survey results, spectropolarimetry is capable
of revealing some surprises among individual objects.  In the last dataset
for our project, we observed the well-known Seyfert~2 galaxy NGC~2110.  For
some reason, this object had been passed over in other surveys at H$\alpha$.
As Figure~6 illustrates, we have detected a spectacular double-peaked broad
H$\alpha$ line in the polarized-flux spectrum of this object (Moran et al.\
2007).  The line profile is similar to those of other double-peaked emitters
(Eracleous \& Halpern 2003).  The remarkable thing is that the line profile
has not been smeared out in the scattering process.  The H$\alpha$ emission
originates in a disk, so most of the scattering must be done by a medium that
sees that disk over a pretty narrow range of inclination angles.  Future
spectropolarimetry surveys may uncover similar gems.

\begin{figure}[htb]
\centerline{\includegraphics[scale=0.52]{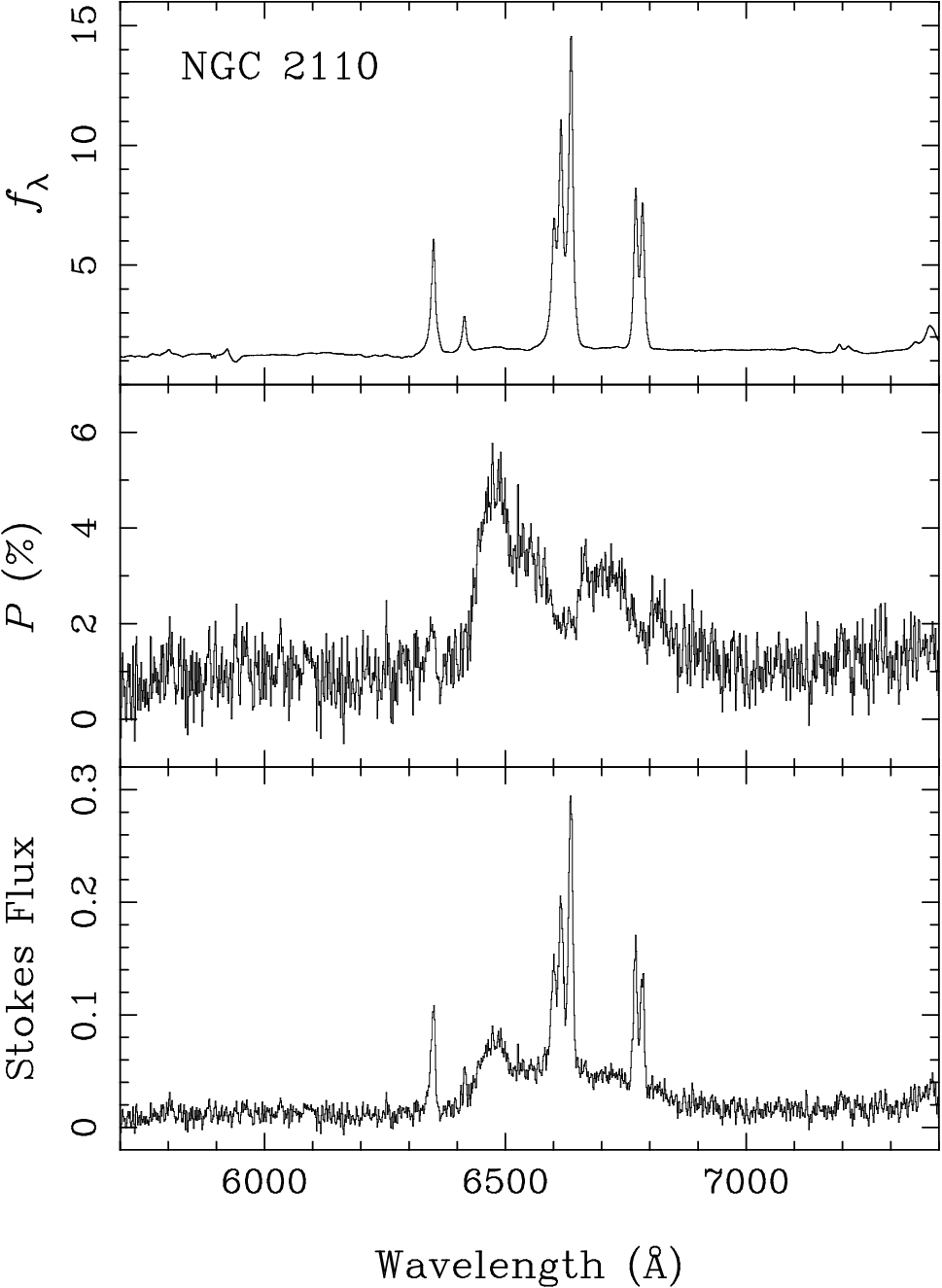}}
\caption{Spectropolarimetry data for NGC 2110.  From top to bottom, the panels
display the total flux, percent polarization, and polarized flux.}
\end{figure}

\acknowledgements
I would like to thank the conference organizers for the opportunity to discuss
the results that have been so important to our current understanding of
physical nature of AGNs.  I am also grateful to my collaborators --- Aaron
Barth, Laura Kay, Alex Filippenko, and Mike Eracleous --- for their extensive
contributions to this work.

\end{document}